%% file: main.tex
\documentclass[10pt,twocolumn,letterpaper]{article}

 \usepackage{cvpr}              
\input{preamble}
\definecolor{cvprblue}{rgb}{0.21,0.49,0.74}
\usepackage[pagebackref,breaklinks,colorlinks,allcolors=cvprblue]{hyperref}

\usepackage{algorithm}
\usepackage{algpseudocode}

\usepackage{float}

\usepackage{amsmath}
\usepackage{booktabs}
\usepackage{tabularx}
\usepackage{graphicx}
\usepackage{xcolor} 
\usepackage{multirow}

\def\paperID{17317} 
\def\confName{CVPR}
\def\confYear{2026}

\title{Multi-Objective Reinforcement Learning for Large-Scale Mixed Traffic Control}

\author{
Iftekharul Islam \quad
Weizi Li \\
University of Tennessee, Knoxville, USA\\
{\tt\small mislam73@vols.utk.edu, weizili@utk.edu}
}

\begin{document}
\maketitle

\input{sec/0_abstract}

\input{sec/i_intro}
\input{sec/ii_related_work}

\input{sec/iii_methodology}

\input{sec/iv_results}
\input{sec/v_conclusion}

{
    \small
    \bibliographystyle{ieeenat_fullname}
    \bibliography{main}
}


\end{document}

%% file: sec/0_abstract.tex

\begin{abstract}
Effective mixed traffic control requires balancing efficiency, fairness, and safety. Existing approaches excel at optimizing efficiency and enforcing safety constraints but lack mechanisms to ensure equitable service, resulting in systematic starvation of vehicles on low-demand approaches. We propose a hierarchical framework combining multi-objective reinforcement learning for local intersection control with strategic routing for network-level coordination. Our approach introduces a Conflict Threat Vector that provides agents with explicit risk signals for proactive conflict avoidance, and a queue parity penalty that ensures equitable service across all traffic streams. Extensive experiments on a real-world network across different robot vehicle (RV) penetration rates demonstrate substantial improvements: up to 53\% reductions in average wait time, up to 86\% reductions in maximum starvation, and up to 86\% reduction in conflict rate compared to baselines, while maintaining fuel efficiency. Our analysis reveals that strategic routing effectiveness scales with RV penetration, becoming increasingly valuable at higher autonomy levels. The results demonstrate that multi-objective optimization through well-curated reward functions paired with strategic RV routing yields significant benefits in fairness and safety metrics critical for equitable mixed-autonomy deployment.
\end{abstract}

%% file: sec/i_intro.tex
\section{Introduction}
\label{sec:introduction}


Urban traffic congestion imposes a staggering economic burden on modern metropolitan areas. American commuters collectively waste over 4 billion hours in traffic delays annually, with the average U.S. driver losing 43 hours—approximately one full work week—to these holdups~\cite{CNBC2025INRIX}. The total national economic losses from this congestion exceeds \$74 billion, representing lost time and productivity for drivers and businesses~\cite{INRIX2024Scorecard}. Beyond direct economic costs, such bottlenecks contribute to increased emissions, wasted fuel, and reduced quality of life, underscoring the critical need for more intelligent and adaptive traffic management systems~\cite{lu2021expansion}.


Recent advances in connected and autonomous vehicle technologies have created new opportunities for congestion mitigation through intelligent, data-driven control. Yet the transition to full autonomy will be incremental, ushering in a mixed-traffic era where Robot Vehicles (RVs) and Human-driven Vehicles (HVs) must coexist and interact within large urban networks~\cite{wu2017flow, peng2021connected}. Managing such systems is challenging due to the unpredictability of HV behavior and the uneven spatial distribution of RVs, which can destabilize coordinated flow under dynamic demand. Reinforcement Learning (RL) has shown strong potential for addressing these challenges by enabling RVs to learn decentralized, adaptive control policies that respond to real-time conditions, outperforming traditional traffic signals at unsignalized intersections and small-scale networks~\cite{Pan2025Review,Liu2025Large,Islam2025Heterogeneous,Fan2025OD,Wang2024Intersection,Poudel2024CARL,Poudel2024EnduRL,Villarreal2024Eco,Villarreal2023Pixel,Villarreal2023Chat, yan2021reinforcement, al2023self}.

However, existing efforts remain limited in two important ways. First, most intersection-level policies employ reactive safety mechanisms that penalize collisions after they occur, rather than enabling proactive conflict avoidance through anticipatory risk awareness. Furthermore, these policies lack explicit fairness guarantees, leading to systematic starvation of vehicles on low-demand approaches even as overall efficiency improves. Second, when scaled to network-level control, these methods encounter the RV shortage problem: imbalanced RV distribution across the network can leave some intersections entirely uncoordinated, negating the benefits of autonomy~\cite{Wang2024Privacy}.

In this study, we aim to fill this critical gap by proposing a hierarchical framework that integrates multi-objective RL with strategic network-level coordination. We conduct comprehensive experiments to evaluate efficiency, fairness, safety, and sustainability of mixed traffic control across varying RV penetration rates. Specifically, our contributions are the following:

\begin{itemize}[leftmargin=*]
    \item We introduce a \textbf{Conflict Threat Vector} that provides agents with explicit, pre-computed risk signals for proactive collision avoidance, moving beyond purely reactive safety mechanisms. This enables agents to learn inherently cautious behaviors that avoid conflict-prone situations before they escalate.
    
    \item We design a \textbf{multi-objective reward function} that explicitly balances efficiency, fairness, and proactive safety, enabling policies that service all traffic streams equitably without sacrificing overall performance.
    
    \item We propose a \textbf{hierarchical framework} combining tactical RL control at intersections with strategic routing for network-level RV distribution, demonstrating that coordination effectiveness scales with RV penetration rates.
    
    \item Our experiments on an 18-intersection real-world network demonstrate substantial improvements: 20--53\% reductions in average wait time, 60--86\% reductions in maximum starvation, and up to 86\% reduction in conflict rate compared to baselines, while maintaining comparable fuel efficiency.

\end{itemize}

%% file: sec/ii_related_work.tex
\section{Related Work}
\label{sec:related_work}

The challenge of coordinating mixed traffic has been increasingly addressed by multi-agent reinforcement learning (MARL)~\cite{Poudel2024CARL, villarreal2023mixed, al2023self}. This approach has scaled from simple environments~\cite{yan2021reinforcement, peng2021connected} to complex, large-scale networks~\cite{wang2023learning, Liu2025Large}. However, existing policies often rely on reward functions that prioritize throughput and enforce safety via reactive, post-collision penalties~\cite{guo2024heuristic, wang2023learning}. This approach is inefficient and provides no explicit guarantee of fairness, often starving low-demand lanes. This has spurred research into multi-objective RL. Some studies focus on fairness via centralized, courteous agents~\cite{yan2021courteous}, while others target proactive safety using predictive classifiers~\cite{Poudel2024EnduRL} or hard-coded, heuristic-based rules like action masks~\cite{feng2025right, guo2024heuristic}. While effective, these solutions often address safety and fairness in isolation. Moreover, heuristic-based methods can be rigid, and centralized controllers lack scalability~\cite{feng2025right, yan2021courteous}.

At the network level, 
research has primarily focused on congestion-aware routing. This field is well-established, with methods ranging from classic graph-search algorithms~\cite{karur2021survey} to advanced systems that use crowd-sourced data to find optimal, time-dependent paths~\cite{bast2016route, abdelrahman2020crowdsensing}. Most of these systems, however, are \textit{ego-centric}, designed to find the best path for a single user~\cite{zhang2022route}. More advanced, socially-aware systems aim for a \textit{system-centric} optimum by distributing vehicles to minimize total network congestion~\cite{wilkie2011self, wilkie2014participatory}.  A distinct problem in mixed traffic is the RV shortage issue. Wang et al.~\cite{Wang2024Privacy} addressed this with a centralized framework using an interactive, multi-step protocol to assign routes. This provides a high degree of central control but also introduces communication and computational overhead.
Our approach features both a decentralized, multi-objective tactical agent with \textit{proactive} safety and \textit{explicit} fairness, as well as a lightweight, \textit{coverage-aware} strategic router, bridging the gaps in both domains.



%% file: sec/iii_methodology.tex
\section{Methodology}
\label{sec:methodology}

We develop a hierarchical control framework that integrates multiple objectives through two synergistic layers: a \emph{control layer}, where robot vehicles (RVs) learn multi-objective decision-making through deep reinforcement learning (RL), and a \emph{routing layer}, which proactively regulates network-wide RV distribution through a lightweight, coverage-aware coordination process. Together, these layers form an integrated system that adapts to both the dynamics of local interactions and the evolving global traffic state.




\subsection{Control Layer: Multi-Objective RL}

We formulate the problem of controlling mixed traffic as a Partially Observable Markov Decision Process (POMDP), defined by the tuple $(\mathcal{S}, \mathcal{A}, \mathcal{P}, \mathcal{R}, \mathcal{O}, \mathcal{Z}, \gamma)$. Here, $\mathcal{S}$ is the set of all possible environment states; $\mathcal{A}$ is the set of all possible actions; $\mathcal{P}$ is the state transition probability function; $\mathcal{R}$ is the reward function; $\mathcal{O}$ is the local observation space for each agent; $\mathcal{Z}$ is the observation probability function; and $\gamma$ is the discount factor. Each RV acts as an agent interacting with the environment at discrete time steps. 
The shared policy \(\pi_\theta\) maps observations \(o_t \in \mathcal{O}\) to actions \(a_t \in \mathcal{A}\), aiming to maximize the expected discounted reward
\(\mathbb{E}_{\pi_\theta}\left[\sum_t \gamma^t \mathcal{R}(s_t, a_t)\right]\).
RVs execute decentralized control during simulation while being trained centrally. 

At the local level, each RV controls its entry behavior within unsignalized intersection control zone (30~\textit{m} radius). Outside this zone, both RVs and HVs follow the Intelligent Driver Model (IDM)~\cite{treiber2013traffic}, ensuring realistic longitudinal behavior and safe car-following. Within the control zone, RVs use the learned RL policy to determine action, implicitly guiding the following HVs.

\textbf{Observation Space.} 
Each RV approaching an intersection $j$ from direction $k$ receives a structured observation vector \(o_t\) consisting of traffic cues at the intersection:
\begin{equation}
o_t = \mathbf{q}_t \oplus \mathbf{w}_t \oplus \mathbf{T}_t \oplus \mathbf{G}_t,
\end{equation}
where $\mathbf{q}_t$ and $\mathbf{w}_t$ denote the normalized queue lengths and average waiting times of all incoming approaches ($N{=}8$ directions). 
$\mathbf{G}_t$ denotes intersection occupancy status, and 
$\mathbf{T}_t$ is the \emph{Conflict Threat Vector}. Unlike reactive safety mechanisms that learn from conflict penalties, $\mathbf{T}_t$ provides the agent with explicit, pre-computed risk signals, enabling proactive conflict avoidance.
For each direction $k \in \mathcal{K}$, we first identify the set of conflicting direction $\mathcal{C}(k)$---those directions that would intersect or merge with movement $k$ within the intersection interior. We then compute the raw conflict pressure by aggregating vehicle occupancy in the entry zones of all conflicting paths:

\begin{equation}
  S_t(j,k) \;=\; \sum_{p\in\mathcal{C}(k)} \sum_{c=1}^{C_0} w_c \, G_t(j,p,c),
  \label{eq:conflict_pressure}
\end{equation}

\noindent where $G_t(j,p,c) \in \{0,1,2,\ldots\}$ denotes the occupancy count in the $c$-th cell of path $p$ at time $t$, with cells $c=1,\ldots,C_0$ covering the critical entry zone nearest to the conflict region. 
Intuitively, this formulation captures how many vehicles on conflicting paths are ``about to enter'' the shared intersection space. Then, the threat score $T_{t,k}$ is given by:

\begin{equation}
  T_{t,k} \;=\; \min\!\left(\frac{S_t(j,k)}{Z_j},\,1\right),
  \label{eq:risk_score}
\end{equation}

\noindent where $Z_j$ normalizes across intersections. The complete Conflict Threat Vector thus becomes $\mathbf{T}_t = \langle T_{t,k} \rangle_{k \in \mathcal{K}}$.

\textbf{Action Space.}
The discrete action set is \(\mathcal{A} = \{\texttt{Stop}, \texttt{Go}\}\).
When an RV decides \texttt{Stop}, it decelerates based on its current speed $u$ and distance to intersection $d_{int}$: $a = -u^2/2d_{int}$; when it chooses \texttt{Go}, it accelerates through the intersection using maximum acceleration.

\textbf{Reward Function.}
We design the reward function to train an agent by balancing three objectives: maximizing local efficiency, maintaining intersection-level fairness, and ensuring proactive safety. To achieve this, the base reward $R_{\text{base}}(s_t, a_t)$ is formulated as a linear combination of three main components:

\begin{equation}
  R_{\text{base}}(s_t, a_t) \;=\; r_{\text{ego}} \;-\; \lambda_{\text{parity}} \cdot r_{\text{parity}} \;-\; \lambda_{\text{threat}} \cdot r_{\text{threat}},
  \label{eq:reward_base}
\end{equation}

where $\lambda_{\text{parity}}$ and $\lambda_{\text{threat}}$ are weighting hyperparameters. The components are:
\begin{enumerate}
    \item \textbf{Egocentric Reward ($r_{\text{ego}}$).} The primary efficiency incentive, based on the normalized waiting time of the agent's own queue $w_{\text{ego}}$. It is positive for a \texttt{Go} action and negative for a \texttt{Stop} action, driving the agent to clear its own queue effectively.
    \item \textbf{Queue Parity Penalty ($r_{\text{parity}}$).} To promote fairness, the agent is penalized by the variance of normalized queue lengths ($\sigma^2_{\mathbf{q}}$) across all $N{=}8$ approaches. A purely egocentric agent might starve a minor traffic stream indefinitely. This penalty incentivizes the agent to learn a more balanced policy that services all approaches, preventing any single queue from starvation.

    \begin{equation}
      r_{\text{parity}} \;=\; \sigma^2_{\mathbf{q}} \;=\; \frac{1}{N} \sum_{i=1}^{N} \big(q_i - \mu_{\mathbf{q}}\big)^2.
      \label{eq:reward_parity}
    \end{equation}

    \item \textbf{Threat Penalty ($r_{\text{threat}}$).} To encourage proactive safety, a \texttt{Go} action is penalized proportionally to its pre-calculated conflict threat score $T_{t,k}$. This provides a dense and continuous reward signal for risk avoidance—effectively a ``yellow light'' that teaches the agent to avoid even the \emph{risk} of a conflict, rather than learning only from actual collisions.
\end{enumerate}

Finally, to ensure absolute safety, a large, discrete negative penalty $p_c = -1$ is applied if a \texttt{Go} action is deemed unsafe and is overridden by the hard-coded conflict resolution mechanism. This acts as a ``red light''—an unambiguous punishment for a critical error. The total reward function $R_{\text{total}}(s_t, a_t)$ is therefore:

\begin{equation}
    R_{\text{total}}(s_t, a_t) \;=\;
    \begin{cases}
    R_{\text{base}}(s_t, a_t) + p_c, & \text{if conflict},\\[2pt]
    R_{\text{base}}(s_t, a_t),       & \text{otherwise}.
    \end{cases}
    \label{eq:reward_total}
\end{equation}

This two-level safety system, combining the proactive threat penalty with the reactive conflict punishment, encourages the agent to learn a policy that is not only efficient and fair but also inherently cautious.

\subsection{Routing Layer}
The routing layer is a privacy-preserving routing algorithm that proactively balances the network-wide distribution of RVs. It consists of a central coordinator and a decentralized policy executed by each RV.

\subsubsection{Proactive Coverage Coordinator}
\label{sec:coverage_coordinator}

The coordinator monitors real-time RV coverage $P_t(e) \in [0,1]$ for each edge $e \in E$, defined as the ratio of RVs to total vehicles on that edge. To enable proactive rebalancing, the system maintains a sliding time-window history:

\begin{equation}
     H_t(e) = \{ P_{t-k+1}(e), \ldots, P_{t-1}(e), P_t(e) \},
    \label{eq:history_window}
\end{equation}

\noindent where $k$ is the history window size. When the history is sufficiently populated ($|H_t(e)| = k$), a linear regression computes the coverage trend $m_t(e)$, allowing prediction of future coverage at horizon $h$:

\begin{equation}
     \hat{P}_{t+h}(e) = \text{clip}\left(P_t(e) + m_t(e) \cdot h, \; 0, \; 1\right).
    \label{eq:coverage_prediction}
\end{equation}

The predicted RV shortage relative to a target coverage $P_{\text{target}}$ is:

\begin{equation}
     \hat{S}_t(e) = \max\big(0,\, P_{\text{target}} - \hat{P}_{t+h}(e)\big).
    \label{eq:shortage_prediction}
\end{equation}

Edges with predicted shortages are made more attractive for routing by reducing their travel costs:

\begin{equation}
     \tau'_t(e) = \tau(e) - \alpha \cdot \hat{S}_t(e) \cdot \tau(e),
    \label{eq:edge_cost_update}
\end{equation}

\noindent where $\tau(e)$ is the baseline travel cost (edge length) and $\alpha$ controls the strength of the routing incentive. The adjusted cost map $\tau'_t$ is broadcast to all RVs at each update interval.

\subsubsection{Decentralized Rerouting Policy}
\label{sec:decentralized_rerouting}

Upon receiving the adjusted cost map $\tau'_t$, each RV autonomously decides whether to compute a new route. To ensure system stability and prevent oscillations, rerouting is subject to several constraints:

\begin{enumerate}[nosep, leftmargin=*]
    \item \textbf{Cooldown period:} An RV cannot reroute if it has rerouted recently, preventing excessive route changes.
    \item \textbf{Commitment distance:} An RV must be sufficiently far from the next junction to avoid last-second route changes that could destabilize traffic flow.
    \item \textbf{Probabilistic activation:} Each eligible RV considers rerouting with probability $\rho \in [0,1]$ to prevent synchronized behavior that could cause oscillations.
\end{enumerate}


If an RV proceeds, it computes a candidate route $R'_{\text{new}}$ by finding the shortest path from its current edge $e_c$ to destination $e_d$ using the adjusted costs $\tau'_t$. Crucially, to prevent inefficient detours, the candidate route must satisfy:

\begin{equation}
     C(R'_{\text{new}}) = \sum_{e \in R'_{\text{new}}} \tau(e) \;\le\; \delta \cdot C(R_{\text{base}}),
    \label{eq:reroute_eligibility}
\end{equation}

\noindent where $C(R_{\text{base}})$ is the cost of the vehicle's original baseline shortest path and $\delta > 1$ is the maximum allowable detour ratio. This ensures that routing for network balance does not excessively increase individual travel times. If the candidate route passes this verification, the RV adopts it locally without reporting back to the coordinator, preserving privacy.
The routing layer is formalized in Algorithms~\ref{alg:coordinator} and~\ref{alg:rv_reroute}.

The routing layer achieves computational efficiency through simple 
operations: $O(E)$ for cost map updates and $O(E log V)$ for shortest path computation. Unlike centralized approaches~\cite{Wang2024Privacy} 
that require iterative vehicle-coordinator communication, our 
broadcast architecture minimizes communication overhead.

\begin{algorithm}[b]
\caption{Proactive Coverage Coordinator}
\label{alg:coordinator}
\begin{algorithmic}[1]
\State \textbf{Input:} Set of edges $E$, current vehicle states
\State \textbf{Output:} Adjusted cost map $\tau'$
\Procedure{GenerateCostMap}{}
    \ForAll{$e \in E$}
        \State Compute current RV coverage $P_t(e)$
        \State Update history $H_t(e) \leftarrow H_{t-1}(e) \cup \{P_t(e)\}$
        \If{$|H_t(e)| = k$}
            \State Compute trend $m_t(e) \leftarrow \textsc{LinearSlope}(H_t(e))$
            \State $\hat{P}_{t+h}(e) \leftarrow \text{clip}(P_t(e) + m_t(e) \cdot h,\, 0,\, 1)$
        \Else
            \State $\hat{P}_{t+h}(e) \leftarrow P_t(e)$
        \EndIf
        \State Predicted shortage: $\hat{S}_t(e) \leftarrow \max(0,\, P_{\text{target}} - \hat{P}_{t+h}(e))$
        \State Adjusted cost: $\tau'_t(e) \leftarrow \tau(e) - \alpha \cdot \hat{S}_t(e) \cdot \tau(e)$
    \EndFor
    \State Broadcast $\tau'_t$ to all RVs
\EndProcedure
\end{algorithmic}
\end{algorithm}

\begin{algorithm}[b]
\caption{Decentralized Rerouting Policy (per RV $v$)}
\label{alg:rv_reroute}
\begin{algorithmic}[1]
\State \textbf{Input:} RV state $(e_c, e_d, C(R_{\text{base}}))$, cost map $\tau'$
\Procedure{ConsiderReroute}{}
    \If{\textbf{not} \textsc{IsEligible}($v$)} \Comment{Check cooldown, distance, etc.}
        \State \Return
    \EndIf
    \If{$\text{random}() > \rho$} \Comment{Probabilistic gating}
        \State \Return
    \EndIf
    \State $R'_{\text{new}} \leftarrow \textsc{ShortestPath}(G, e_c, e_d, \text{weights}=\tau')$
    \State $C(R'_{\text{new}}) \leftarrow \sum_{e \in R'_{\text{new}}} \tau(e)$ \Comment{Verify using baseline costs}
    \If{$C(R'_{\text{new}}) \le \delta \cdot C(R_{\text{base}})$}
        \State $v.\text{route} \leftarrow R'_{\text{new}}$ \Comment{Adopt new route}
        \State \textsc{StartCooldown}($v$)
    \EndIf
\EndProcedure
\end{algorithmic}
\end{algorithm}

\subsection{Evaluation Metrics}
\label{sec:evaluation_metrics}

We assess system performance across four dimensions: \textit{efficiency}, \textit{fairness}, \textit{safety}, and \textit{sustainability}. 

\subsubsection{Efficiency Metrics}

\textbf{Average Wait Time ($W_{\text{avg}}$).} 
This metric reflects the mean duration vehicles remain stationary within intersection control zones~\cite{wang2023learning}. 
For each vehicle, we accumulate the total stopped time; the average is then taken across all vehicles in the network. 
Lower wait times indicate smoother traffic flow and reduced congestion.

\textbf{Throughput ($\Theta$).} 
We measure throughput at two levels: (1) \textit{intersection throughput}, the number of vehicles that successfully cross an intersection within 500~\textit{s}, averaged across all intersections; and (2) \textit{network throughput}, the number of vehicles that complete their trips during the same period. 
Higher throughput values signify greater traffic efficiency and system capacity utilization.

\textbf{Average Delay ($D_{\text{avg}}$).} 
Average delay quantifies the extra travel time experienced compared to free-flow conditions. 
It captures the overall network impact of congestion and control performance.

\subsubsection{Fairness Metrics}

\textbf{Maximum Starvation ($W_{\text{max}}$).} 
Maximum starvation measures the longest consecutive 
period (in seconds) during which any traffic approach maintains average wait times exceeding 60~\textit{s}, indicating sustained starvation of that approach. High values indicate that some approaches experience prolonged neglect under the control policy~\cite{chen2016cooperative}.

\textbf{99th Percentile Wait ($W_{p99}$).} 
This metric captures the tail of the wait time distribution and provides a more robust fairness measure less sensitive to extreme outliers.

\subsubsection{Safety Metric}

\textbf{Conflict Rate ($C_{\text{rate}}$).} 
This measures the proportion of RV \texttt{Go} actions overridden by the built-in safety mechanism due to potential collision risk~\cite{wang2023learning}. 
A lower conflict rate indicates safer, more anticipatory decision-making by the learned policy. 

\subsubsection{Sustainability Metric}

\textbf{Fuel Consumption ($F_{\text{avg}}$).} 
We record the average fuel consumption of vehicles passing through intersection zones using SUMO’s HBEFA3 emission model~\cite{sumo_doc2, sumo_doc3}. 
This metric captures energy efficiency under stop-and-go conditions, reflecting the environmental impact of each control strategy.

%% file: sec/iv_results.tex
\section{Experiments and Results}
\label{sec:experiments}

We evaluate our approach on a real-world urban network under varying RV rates, comparing against traditional traffic light control and a state-of-the-art RL baseline.

\subsection{Experimental Setup}
\label{sec:experimental_setup}

\begin{figure*}[t]
  \centering
  \includegraphics[width=0.95\textwidth]{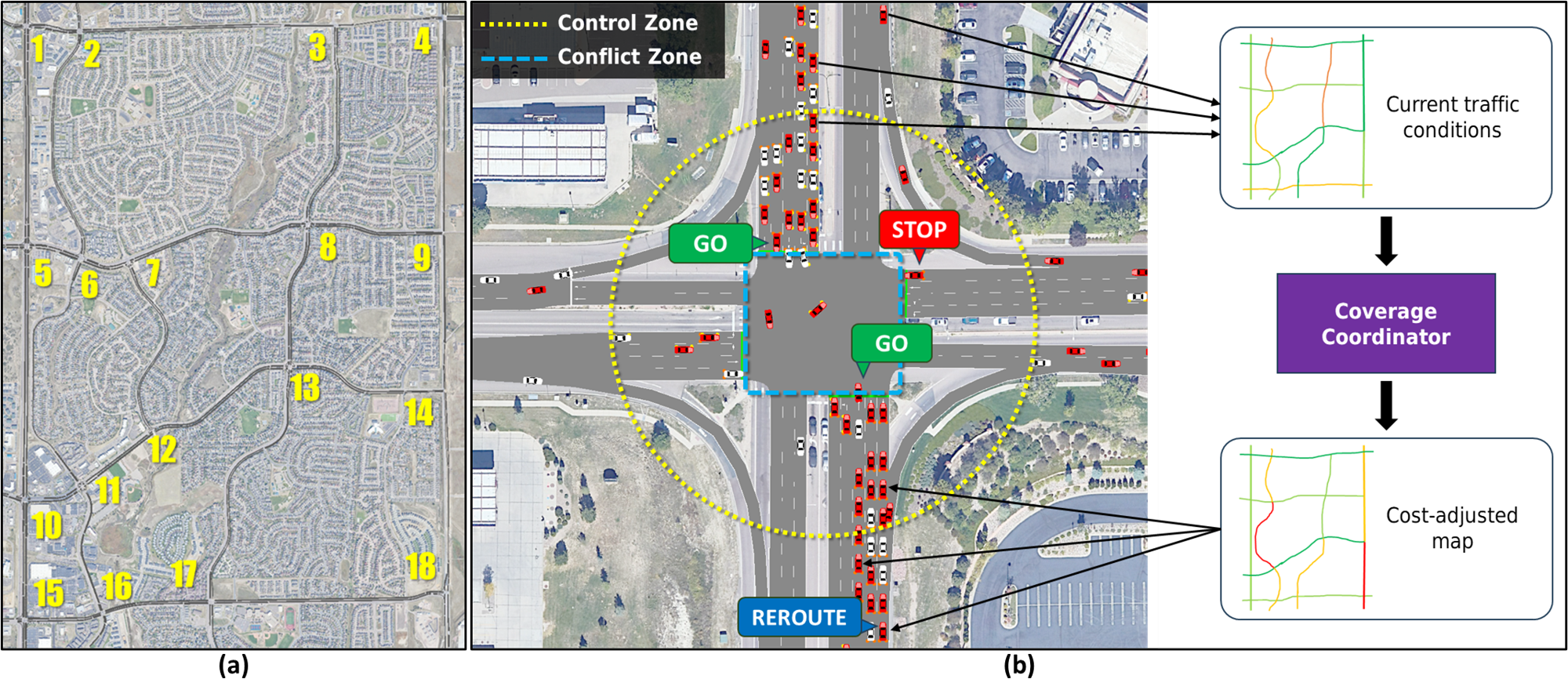}

    \caption{Hierarchical framework combining tactical control and strategic routing. \textbf{(a)} We evaluate our framework on a large-scale, real-world network from Colorado Springs, CO, USA, featuring 18 intersections. \textbf{(b)} System architecture (RVs in red, HVs in white). At intersections, RVs within the control zone make Go/Stop decisions using multi-objective rewards balancing efficiency, fairness, and proactive safety. At the network level, the Coverage Coordinator broadcasts cost-adjusted maps based on traffic conditions and predicted coverage shortages; RVs make decentralized rerouting decisions balancing local efficiency with network-wide coverage.} 
  \label{fig:colorado_network}
\end{figure*}

We conduct experiments on a real-world road network from Colorado Springs, CO, USA, comprising 18 intersections with diverse geometries (Figure~\ref{fig:colorado_network}). Each simulation runs for 1000 \textit{s}, with metrics computed over the steady-state window (500--1000 \textit{s}) to exclude warm-up transients. We evaluate performance across six RV penetration rates (40--90\% in 10\% increments) to assess scalability from early adoption to near-complete autonomy, conducting 10 independent runs per configuration. Each newly spawned vehicle is randomly assigned as an RV or HV according to the specified RV penetration rate.


We compare four methods: \textbf{HV-Sig} (0\% RV, traffic signals with IDM~\cite{treiber2013traffic}), \textbf{Wang et al.}~\cite{wang2023learning} (state-of-the-art RL for intersection control), \textbf{OurRL} (our multi-objective RL with Conflict Threat Vector and queue parity penalty), and \textbf{OurRL+R} (OurRL with strategic routing).


We train shared policies using Rainbow DQN~\cite{hessel2018rainbow} for 1,000 iterations per RV penetration rate on an Intel i9-13900KF CPU with NVIDIA RTX 4090 GPU (30--40 hours per configuration). Hyperparameters were selected through preliminary experiments to balance efficiency, fairness, and safety objectives. 
Table~\ref{tab:hyperparameters} provides complete specifications.

\begin{table}[t]
\centering
\caption{Experimental configuration and hyperparameters.}
\label{tab:hyperparameters}
\small
\begin{tabular}{@{}ll@{}}
\toprule
\textbf{Parameter} & \textbf{Value} \\
\midrule
\multicolumn{2}{l}{\textit{RL Training (Rainbow DQN)}} \\
Architecture & 3 layers, 512 units, ReLU \\
Learning rate, $\gamma$ & $5 \times 10^{-4}$, 0.99 \\
Training iterations & 1,000 \\
Hardware & i9-13900KF, RTX 4090 \\
\midrule
\multicolumn{2}{l}{\textit{Reward Function}} \\
$\lambda_{\text{parity}}$, $\lambda_{\text{threat}}$, $p_c$ & 0.2, 0.5, -1 \\
\midrule
\multicolumn{2}{l}{\textit{Conflict Threat Vector}} \\
$C_0$, $w_c$, $Z_j$ & 3, 1 (uniform), 5 \\
\midrule
\multicolumn{2}{l}{\textit{Strategic Routing}} \\
$\rho$, $\delta$, $\alpha$ & 0.15, 1.20, 1.0 \\
Commitment dist., cooldown & 50 m, 60 steps \\
$P_{\text{target}}$, update interval, $k$ & RV rate - 5\%, 60 steps, 5 \\
\bottomrule
\end{tabular}
\end{table}

\subsection{Performance Analysis}
\label{sec:results}

We present results organized to highlight the impact of our approach---multi-objective rewards for fairness and safety paired with strategic RV routing---followed by their consequences for overall network performance.

\subsubsection{Fairness and Safety}
\label{sec:results_fairness_safety}

Tables~\ref{tab:fairness} and~\ref{tab:safety} present fairness and safety metrics, revealing the most significant advantages of our multi-objective RL.

\textbf{Fairness.}
As Table~\ref{tab:fairness} shows, 
the HV-Sig baseline exhibits severe starvation, with at least one approach maintaining average wait times above 60 \textit{s} throughout the entire 500-second measurement window. This indicates complete gridlock where vehicles on certain approaches are effectively unable to clear the intersection under fixed-signal control.
The Wang et al. baseline, which lacks an explicit fairness term, improves this but still suffers from highly variable starvation times (270--437 \textit{s} depending on RV rate). OurRL policy, by directly penalizing queue variance, dramatically cuts maximum starvation time. At 90\% RV, OurRL achieves $W_{\text{max}}$ of just 139.5 \textit{s}---a 72\% reduction compared to HV-Sig and 63\% reduction compared to Wang et al. This confirms the agent has learned to service all approaches equitably rather than optimizing purely for throughput. The full OurRL+R framework pushes this even further: by strategically routing RVs to underserved, high-queue areas, the hierarchical system achieves a remarkable $W_{\text{max}}$ of only 69.7 \textit{s} at 90\% RV, an 86\% improvement vs. HV-Sig and 81\% vs. Wang et al. baseline.

\begin{table}[t]
\centering
\caption{Fairness metrics across methods and RV penetration rates. OurRL achieves up to 63\% reductions in maximum starvation vs. Wang et al. and up to 72\% vs. HV-Sig through the queue parity penalty. With routing at 90\% RV, starvation drops to 69.7 \textit{s} (86\% improvement over HV-Sig).}
\label{tab:fairness}
\small
\begin{tabular}{@{}lrrrrrr@{}}
\toprule
\multirow{2}{*}{\textbf{Method}} & \multicolumn{6}{c}{\textbf{RV Penetration Rate}} \\
\cmidrule(l){2-7}
 & \textbf{40\%} & \textbf{50\%} & \textbf{60\%} & \textbf{70\%} & \textbf{80\%} & \textbf{90\%} \\
\midrule
\multicolumn{7}{l}{\textit{Maximum Starvation Time (\textit{s}, lower is better)}} \\
HV-Sig (0\%) & \multicolumn{6}{c}{500.0} \\
Wang et al. & 437.1 & 344.2 & 269.9 & 272.1 & 330.1 & 374.8 \\
OurRL & \textbf{242.5} & 254.9 & 228.7 & 196.2 & 191.8 & 139.5 \\
OurRL+R & 281.0 & \textbf{210.6} & \textbf{177.3} & \textbf{108.6} & \textbf{151.9} & \textbf{69.7} \\
\midrule
\multicolumn{7}{l}{\textit{99th Percentile Wait Time (\textit{s}, lower is better)}} \\
HV-Sig (0\%) & \multicolumn{6}{c}{135.0} \\
Wang et al. & 151.1 & 127.6 & 142.9 & 87.5 & 101.6 & 128.2 \\
OurRL & \textbf{106.7} & 103.1 & 94.7 & 70.7 & 84.2 & 60.6 \\
OurRL+R & 126.1 & \textbf{95.9} & \textbf{70.1} & \textbf{55.3} & \textbf{64.2} & \textbf{48.0} \\
\bottomrule
\end{tabular}
\end{table}

The tail of the wait time distribution shows similar patterns. OurRL achieves 60--107 \textit{s} for $W_{p99}$ across penetration rates, compared to 135 \textit{s} for HV-Sig and 87--151 \textit{s} for Wang et al. With routing enabled at high penetration rates, we achieve $W_{p99}$ as low as 48 \textit{s} (90\% RV), demonstrating that our approach benefits not just average cases but also vehicles experiencing the longest delays.

Interestingly, this fairness-aware approach does not come at the expense of overall efficiency. Our method maintains competitive or superior average wait times compared to the efficiency-focused Wang et al. baseline while dramatically improving fairness (see Section~\ref{sec:results_efficiency}). This suggests that the fairness penalty helps the agent discover policies that are not just efficient on average but robust across diverse traffic patterns.


\textbf{Proactive Safety.}
The Conflict Threat Vector provides agents with explicit, pre-computed risk signals that enable anticipatory decision-making. Unlike reactive safety mechanisms that penalize the agent only after a conflict is detected, our approach provides continuous risk signals that shape the policy during learning. This allows the agent to develop inherently cautious behaviors that avoid conflict-prone situations before they escalate.




Results in Table~\ref{tab:safety} demonstrate the effectiveness of this proactive approach. At 40\% RV, our method achieves 15.69\% conflict rate compared to Wang et al.'s 16.52\%---a modest 5\% relative improvement. However, the performance gap widens dramatically at higher RV penetration rates. At 90\% RV, our method achieves just 2.85\% conflict rate compared to Wang et al.'s 20.58\%---an 86\% relative reduction in conflicts. This increasing effectiveness suggests a positive feedback effect: as more vehicles operate under coordinated RL control with shared risk awareness, the overall predictability of intersection behavior improves, enabling even safer coordination. The routing layer reinforces this trend, maintaining conflict rates below 6\% at all RV rates of 60\% or higher. 

The data reveal distinct performance regimes by RV density. At lower penetration rates (40--50\%), conflict rates show higher variability (15.69--23.55\%), reflecting the challenge of coordinating mixed traffic when RV presence is sparse and HV behavior dominates. At 60\% RV and above, conflict rates drop sharply and remain consistently low (8.32\% to 2.62\%) as sufficient RV density enables effective coordinated control. This trend bodes well for real-world deployment scenarios where AV adoption will gradually increase over time.

\begin{table}[t]
\centering
\caption{Conflict rate (\%) for RL-based methods across RV penetration rates. Our Conflict Threat Vector achieves up to 86\% conflict reduction at 90\% RV (2.85\% vs. 20.58\% for Wang et al.), with increasing effectiveness at higher penetration rates.}
\label{tab:safety}
\small
\begin{tabular}{@{}lrrrrrr@{}}
\toprule
\multirow{2}{*}{\textbf{Method}} & \multicolumn{6}{c}{\textbf{RV Penetration Rate}} \\
\cmidrule(l){2-7}
 & \textbf{40\%} & \textbf{50\%} & \textbf{60\%} & \textbf{70\%} & \textbf{80\%} & \textbf{90\%} \\
\midrule
Wang et al. & 16.52 & 23.97 & 22.09 & 17.42 & 15.32 & 20.58 \\
OurRL & \textbf{15.69} & \textbf{20.67} & 8.32 & \textbf{4.56} & 7.31 & 2.85 \\
OurRL+R & 19.97 & 23.55 & \textbf{5.57} & 4.58 & \textbf{5.84} & \textbf{2.62} \\
\bottomrule
\end{tabular}
\end{table}

\subsubsection{Overall Efficiency}
\label{sec:results_efficiency}

\begin{figure*}[ht]
  \centering
   \includegraphics[width=1.0\linewidth]{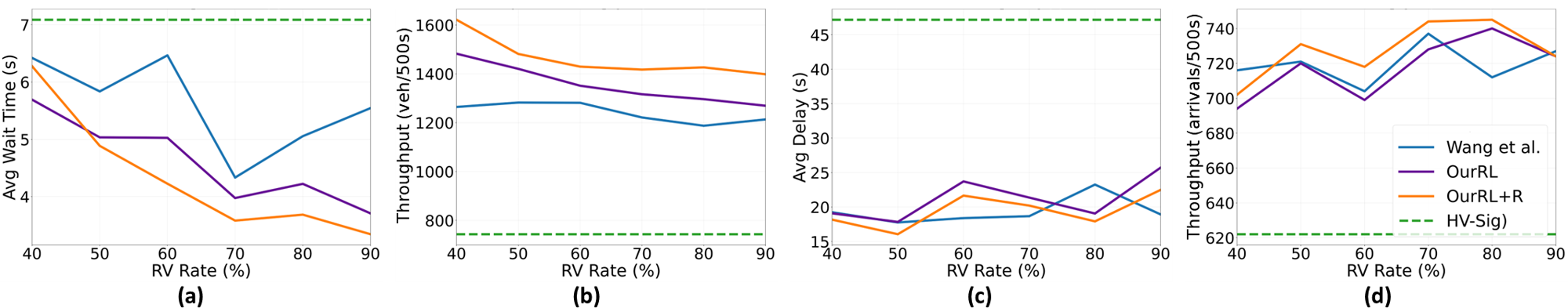}
   \caption{Efficiency metrics across methods and RV penetration rates. (a) Average wait time: OurRL achieves 11--40\% improvements over Wang et al. and 20--53\% over HV-Sig, with OurRL+R reaching 3.34~\textit{s} at 90\% RV. (b) Intersection throughput: OurRL achieves 1270--1483 \textit{veh/500s} per intersection, with routing providing additional local gains (1398--1621 \textit{veh/500s}). (c) Average delay: RL methods reduce delay from 47.16~\textit{s} (HV-Sig) to 16--26~\textit{s} range. (d) Network throughput: All RL methods achieve 690--745 \textit{veh/500s} (11--20\% improvement) compared to HV-Sig's 622 \textit{veh/500s}.}
   \label{fig:efficiency}
\end{figure*}

Figure~\ref{fig:efficiency} presents efficiency metrics across all methods and RV penetration rates. A key finding emerges: by preventing starvation and conflicts (as shown in Section~\ref{sec:results_fairness_safety}), our multi-objective policy achieves superior overall network performance compared to 
the baseline methods.

\textbf{Local (Intersection-Level) Efficiency.}
In terms of 
\textit{average wait time}, OurRL policy consistently outperforms both baselines across all RV penetration rates. At 40\% RV penetration, our method achieves 5.69 \textit{s} average wait time compared to 7.09 \textit{s} for HV-Sig (20\% improvement) and 6.42 \textit{s} for Wang et al. (11\% improvement). The performance gap widens at higher penetration rates: at 90\% RV, our method achieves 3.71 \textit{s}---a 48\% improvement over HV-Sig and 33\% improvement over Wang et al.
These efficiency gains are a direct consequence of the fairness and safety mechanisms. By preventing systematic starvation (through queue parity) and reducing conflicts (through proactive threat awareness), the agent avoids local gridlock conditions that would otherwise propagate through the network. The strategic routing layer builds on this foundation, providing additional gains at higher RV rates by balancing RV distribution across the network (Section~\ref{sec:results_routing}). The result is not just fairer and safer traffic flow, but more efficient performance.

\textit{Intersection throughput} reveals patterns that complement the wait time results. OurRL achieves 1270--1483 vehicles per 500 seconds (\textit{veh/500s}) per intersection across RV rates, consistently outperforming Wang et al. (1188--1282 \textit{veh/500s}). The routing layer provides additional gains, achieving 1398--1621 \textit{veh/500s} at different penetration rates. This demonstrates that our tactical policy is effective at the intersection level, while routing can further enhance processing capacity by directing RVs to under-covered areas. 

\textbf{Network-Wide Efficiency.}
All RL-based methods substantially improve \textit{network throughput} compared to HV-Sig. The traffic light baseline achieves 622~\textit{veh/500s}, while RL methods consistently achieve 690--745~\textit{veh/500s}---representing 11--20\% improvements. OurRL shows competitive throughput with Wang et al., and the routing layer provides additional gains at higher penetration rates (745 \textit{veh/500s} at 80\% RV with routing). Network throughput shows some variability across RV rates, likely due to 
stochastic traffic patterns. However, all RL methods consistently outperform HV-Sig, showing robust efficiency gains.


Similarly, all RL methods dramatically reduce average \textit{delay} compared to HV-Sig (47.16~\textit{s}), achieving delays in the 16--26 \textit{s} range. Our method with routing achieves competitive delays across all penetration rates, with particularly strong performance at 50\% (16.05~\textit{s}) and 80\% RV (17.91~\textit{s}). The delay improvement reflects the combined benefits of reduced conflicts, improved fairness, and strategic routing at the system level.

\subsubsection{Impact of Strategic Routing} \label{sec:results_routing} The strategic routing layer demonstrates penetration-rate-dependent effectiveness, with a notable tradeoff between throughput and wait time at lower RV densities. 

At 40\% RV, routing exhibits an interesting tradeoff. Average intersection throughput improves from 1483 to 1621~\textit{veh/500s} (9\% gain), indicating that routing successfully enhances system capacity by directing RVs toward under-covered intersections. However, average wait time slightly increases from 5.69~\textit{s} to 5.85~\textit{s}, indicating that individual vehicles experience longer delays. This pattern reflects a fundamental challenge at low RV density: redistributing the limited pool of controllable vehicles improves processing capacity at some intersections but creates RV-depleted zones at others. The improved capacity is offset by degraded service quality, with the net effect being longer average wait times despite higher throughput. This suggests that at low RV density, the system can process more vehicles but at the cost of longer individual delays. 

In contrast, at higher RV rates, routing provides substantial improvements across both metrics. At 90\% RV with routing, we observe the best performance across nearly all metrics: 3.34 \textit{s} average wait time (40\% improvement over Wang et al.), 69.7 \textit{s} maximum starvation (81\% improvement over Wang et al.), 48 \textit{s} $W_{p99}$, and 2.62\% conflict rate. Intersection throughput remains high (1398 \textit{veh/500s}) while network wait time decreases. These findings reveal that routing effectiveness depends 
strongly on RV density: below approximately 60\% RV, redistribution 
can create local imbalances, but beyond this threshold, sufficient 
controllable vehicles enable cooperative coverage that enhances 
both fairness and efficiency without creating new shortage zones.



\subsubsection{Sustainability}
\label{sec:results_sustainability}

Fuel consumption, as shown in Table~\ref{tab:sustainability}, demonstrates the sustainability benefits of RL-based coordination. All RL methods achieve 749--763 \textit{ml/s} compared to HV-Sig's 798.3 \textit{ml/s}---a 6\% reduction. This improvement stems from better coordination that reduces overall idling time and enables smoother traffic progression. Notably, fuel consumption remains consistent across all RL methods (749--763 \textit{ml/s}) despite their substantial differences in fairness and safety performance. Our method achieves dramatic fairness improvements (60--86\% starvation reduction) and safety gains (up to 86\% conflict reduction) while maintaining fuel consumption comparable to Wang et al. 
This demonstrates that our multi-objective policy improves traffic progression through smarter coordination---servicing all approaches equitably and avoiding conflicts---without inducing energy-wasteful driving behaviors. The fairness and safety benefits come from better policy decisions rather than from changes in vehicle energy consumption.



\begin{table}[t]
\centering
\caption{Fuel consumption (\textit{ml/s}) across methods and RV penetration rates. Fuel consumption remains consistent (749--763 \textit{ml/s}) across all RL methods despite substantial efficiency and fairness gains, indicating energy-efficient traffic progression.}
\label{tab:sustainability}
\small
\begin{tabular}{@{}lrrrrrr@{}}
\toprule
\multirow{2}{*}{\textbf{Method}} & \multicolumn{6}{c}{\textbf{RV Penetration Rate}} \\
\cmidrule(l){2-7}
 & \textbf{40\%} & \textbf{50\%} & \textbf{60\%} & \textbf{70\%} & \textbf{80\%} & \textbf{90\%} \\
\midrule

HV-Sig (0\%) & \multicolumn{6}{c}{798.3} \\
Wang et al. & 749.2 & 749.5 & 757.1 & 752.3 & 760.1 & 755.9 \\
OurRL & 756.4 & 753.5 & 761.2 & 759.0 & 757.2 & 763.0 \\
OurRL+R & 758.7 & 753.6 & 756.6 & 754.4 & 756.4 & 759.6 \\
\bottomrule
\end{tabular}
\end{table}

\subsection{Summary of Key Findings}
\label{sec:summary}

Our experimental evaluation yields several key insights:

\begin{enumerate}[nosep, leftmargin=*]
    \item \textbf{Multi-objective rewards enable dramatic fairness and safety gains:} The queue parity penalty achieves up to 86\% reductions in maximum starvation compared to baselines, while the Conflict Threat Vector achieves up to 86\% reduction in conflict rate at high RV rates.
    
    \item \textbf{Fairness and safety drive efficiency:} By preventing starvation and conflicts, our multi-objective policy achieves 11--40\% improvements in average wait time over Wang et al. and 20--53\% over HV-Sig, demonstrating that multi-objective optimization yields superior overall network performance.
    
    \item \textbf{Hierarchical coordination scales with penetration:} Strategic routing exhibits a minimum RV density threshold (between 50--60\%) below which local gains come at network-level costs. At higher penetration rates (70--90\%), routing provides substantial synergistic benefits.
    
    \item \textbf{Sustainability without compromise:} Efficiency and fairness improvements are achieved while maintaining fuel consumption comparable to baselines, indicating energy-efficient traffic progression.
\end{enumerate}


%% file: sec/v_conclusion.tex
\section{Conclusion}
\label{sec:conclusion}

This work presents a hierarchical framework for mixed traffic control that balances efficiency, fairness, and proactive safety through multi-objective reinforcement learning. Our approach introduces a Conflict Threat Vector for anticipatory conflict avoidance and a queue parity penalty for equitable service across traffic streams. The strategic routing layer provides a lightweight, communication-efficient mechanism for network-level coordination, whose benefits scale naturally with RV penetration. Evaluation on an 18-intersection urban network across diverse RV penetration rates demonstrates substantial improvements: up to 53\% reductions in average wait time, 60--86\% reductions in maximum starvation time, and up to 86\% reduction in conflict rate, while maintaining comparable fuel efficiency. 
These results show that explicit multi-objective optimization through carefully designed reward functions combined with strategic routing yields substantial real-world benefits in fairness and safety metrics critical for equitable mixed-autonomy deployment, while simultaneously improving overall efficiency.

Several directions merit future investigation. 
The routing coordinator's linear predictor could be enhanced with learning-based models to capture complex dynamics, and explicit RV allocation with real-time rerouting feedback could improve coordination precision~\cite{Guo2024Simulation,Chao2020Survey,Li2017CityFlowRecon,Wilkie2015Virtual}. Adaptive strategies that activate routing based on local RV density thresholds could address variable performance at lower penetration rates. Extension to hybrid networks with both signalized and unsignalized intersections would enhance deployment viability, and real-world pilot studies with V2X integration remain essential for addressing practical challenges. Finally, our study is limited to intersection scenarios, excluding other road types such as one-way corridors or roundabouts. Extending future evaluations to more diverse network layouts would improve the generalizability and real-world applicability of the proposed method~\cite{Poudel2025Urban,Raskoti2025MIAT,Lin2022GCGRNN,Lin2019Compress,Li2018CityEstIET,Li2017CitySparseITSM}. 